\documentclass[12pt,amsart,superscriptaddress,amsfonts,floatfix]{iopart}
\usepackage{fullpage,graphicx,epsf}
\usepackage{iopams}
\usepackage{amsfonts,amssymb}
\usepackage{amscd}
\usepackage{cite}
\usepackage{color}
\usepackage{graphicx}
\usepackage{epsfig}
\usepackage[english]{babel}
\usepackage{amsfonts}
\usepackage{soul}
\usepackage[mathscr]{euscript}
\begin{document}

\def\th{\theta}
\def\l{\label}
\def\p{\partial}
\def\be{\begin{equation}}
\def\ee{\end{equation}}
\def\bea{\begin{eqnarray}}
\def\eea{\end{eqnarray}}
\def\bef{\begin{figure}}
\def\eef{\end{figure}}
\def\bml{\begin{mathletters}}
\def\eml{\end{mathletters}}
\def\l{\label}
\def\b{\beta}
\def\no{\nonumber}
\def\fr{\frac}
\def\o{\omega}
\def\O{\omega}
\def\p{\partial}
\def\n{\nabla}
\def\a{\alpha}
\def\b{\beta}
\def\eps{\epsilon}
\def\g{\gamma}
\def\d{\delta}
\newcommand{\dd}{\mbox{d}}
\title{A stochastic model of long-range interacting particles}
\author{Shamik Gupta$^1$, Thierry Dauxois$^2$, Stefano Ruffo$^3$}
\address{$^1$ Univ. Paris-Sud, CNRS, LPTMS, UMR8626, Orsay
F-91405, France}
\address{$^2$ Laboratoire de Physique de l'\'{E}cole Normale Sup\'{e}rieure de Lyon, Universit\'{e}
de Lyon, CNRS, 46 All\'{e}e d'Italie, 69364 Lyon c\'{e}dex 07, France}
\address{$^3$ Dipartimento di Fisica e Astronomia and CSDC,
Universit\`{a} di Firenze, INFN and CNISM, via G. Sansone, 1 50019 Sesto
Fiorentino, Italy}
\ead{shamikg1@gmail.com,thierry.dauxois@ens-lyon.fr,stefano.ruffo@gmail.com}

\begin{abstract}
We introduce a model of long-range interacting particles evolving under
a stochastic Monte Carlo dynamics, in which possible increase or decrease
in the values of the dynamical variables is accepted with preassigned probabilities. For
symmetric increments, the system at long times settles to the Gibbs
equilibrium state, while for asymmetric updates, the steady state
is out of equilibrium. For the associated Fokker-Planck dynamics in the
thermodynamic limit, we compute exactly the phase space distribution in
the nonequilibrium steady state, and find that it has a nontrivial form
that reduces to the familiar Gibbsian measure in the equilibrium limit.
\end{abstract}
\pacs{05.70.Ln, 02.50.Ey}  
Keywords: stochastic particle dynamics (theory), stationary states
\maketitle
\tableofcontents

\section{Introduction}
Nonequilibrium systems abound in
nature, with examples encompassing different branches of science.
Although there has been much recent progress in characterizing
and understanding some features of nonequilibrium steady states
\cite{Derrida:2011}, developing a general principle akin to the one due to Gibbs-Boltzmann for
equilibrium has been one of the greatest
challenges of modern statistical physics. In this
respect, it is instructive to develop and analyze simple models in order
to gain insights into features of nonequilibrium steady states that make
them distinct from those in equilibrium. Often, even for simple models,
 the steady state distribution has been nontrivial to obtain \cite{Mallick:2006}, and in many
cases, has even remained elusive, thereby
requiring one to resort to numerical simulations and approximation methods as only possible
tools to analyze the steady states \cite{Privman:1997}. 

Here, we develop a model of
particles interacting via long-range interactions and evolving under a stochastic Monte Carlo dynamics, for which we could characterize exactly the steady
state distribution. This is one of the first examples of engineering
such a dynamics in the arena of long-range models
interacting with an external heat bath, where all previous studies, to the best of our knowledge, have been based on
 Langevin equations with noise terms that mimic the effect of the heat
 bath (see, e.g., \cite{Baldovin:2009}). The model effectively simulates driven motion of particles in one dimension under the
 action of a mean-field.    
 
 Long-range interactions have generated
considerable interest in recent years, with examples ranging from plasma physics to gravitational systems \cite{review1}. These systems are characterized
by an interparticle potential which in $d$ dimensions decays
at large separation, $r$, as $1/r^\alpha$, with $\alpha \le d$.
Long-range interacting systems are different from the
short-range ones in that
they are generically non-additive, whereby thermodynamics quantities
scale superlinearly with the system size. This latter feature manifests
in properties, both static and dynamic, which are unusual for
short-range systems \cite{review1}.

In this work, we introduce a model of long-range interacting systems
involving $N$ globally coupled particles moving on a circle, in presence of an external field
acting individually on the particles, and in contact with an external
heat bath at inverse temperature $\beta$. This inverse temperature will
coincide with the steady state temperature of the system only in
equilibrium. The system evolves under a
stochastic Monte Carlo dynamics. Thus, a randomly chosen particle
decides to move either to the left or to the right of its present
position to take up a new location on the circle. The displacement
from the initial position of the particle is a quenched random variable
sampled independently from a common distribution for all the particles. 
The new location of the particle is accepted with a preassigned
transition rate which is chosen in the following way. In the case the
particle jumps symmetrically to the left and to the right, the transition rate is such that the stationary state
of the system is the Gibbs equilibrium state at the inverse temperature
$\beta$. In the case of asymmetric particle jump to the left and
to the right, the system at long times reaches a nonequilibrium
stationary state. Considering the dynamics in the Fokker-Planck limit
(i.e., only small
jumps are allowed), we find that even with asymmetric jumps, if the
external field is turned off and the jump distribution is a delta
function so that all particles jump by the same amount, the steady state is in equilibrium in a suitable
comoving frame obtained by performing a Galilean transformation. In all
other cases, the steady state is out of equilibrium.
In the thermodynamic limit $N \to
\infty$, the system is characterized by a single-particle
distribution, that we compute exactly in the nonequilibrium steady state. We find that the distribution has a non-trivial
form that reduces to the usual Gibbsian distribution in the equilibrium
limit. Our results show excellent agreement with $N$-particle Monte 
Carlo simulations of the dynamics.

The paper is organized as follows. In Section \ref{model}, we give a
precise definition of the model and discuss the Master equation for the
evolution of the $N$-particle phase space distribution. In Section \ref{Langevin}, we
consider the Fokker-Planck limit of the dynamics and obtain the exact
steady state single-particle distribution in the limit $N \to \infty$. In Section
\ref{numerics}, we compare $N$-particle Monte Carlo simulation results
for the steady state single-particle distribution with our theoretical predictions in
the Fokker-Planck limit, and demonstrate an excellent agreement between
the two. The paper ends with conclusions.
\section{Model}
\l{model}
Consider a system of $N$ interacting particles moving on a unit
circle, with the particles labelled as $i=1,2,\ldots,N$. Let the angle
$\th_i$ denote the location of the $i$th
particle on the circle. A microscopic configuration of the
system is denoted by $\mathcal{C}=\{\th_i;i=1,2,\ldots,N\}$. The particles interact via
a long-range potential $\mathcal{V}(\mathcal{C})=K/(2N)\sum_{i,j=1}^N[1-\cos(\th_i-\th_j)]$, where
$K$ is the coupling constant; we consider $K=1$ in the following.
Application of an external field of strength $h_i$ produces a potential $\mathcal{V}_{\rm
ext}(\mathcal{C})=\sum_{i=1}^N h_i \cos \th_i$, so that the net potential energy is
$V(\mathcal{C})=\mathcal{V}(\mathcal{C})+\mathcal{V}_{\rm
ext}(\mathcal{C})$. The interaction $\mathcal{V}(\mathcal{C})$ has the
same form as in the Hamiltonian mean-field (HMF) model, a paradigmatic example of
systems with long-range interactions \cite{review1}. The fields
$h_i$'s may be considered as quenched random variables with a common
distribution $\mathscr{P}(h)$.

We now specify the dynamics of the system. We take hints from one of the first models devoted
to studying characteristics of nonequilibrium steady states, the
celebrated Katz-Lebowitz-Spohn model \cite{KLS}. The configuration $\mathcal{C}$ evolves according to a stochastic
Monte Carlo dynamics. In discrete time, the dynamics in a small time
$\Delta t$ involves every particle attempting to hop to a new position on the
circle. The $i$th particle attempts with probability $p$ to move forward
(in the counter clockwise sense) by an amount $f_i$ on the circle, $\th_i
\to \th_i'=\th_i+f_i$, while with probability $q=1-p$, it
attempts to move backward by the amount $f_i$, so that $\th_i \to
\th_i'=\th_i-f_i$. In either case, the particle takes up the new position with
probability $g(\Delta V(\mathcal{C}))\Delta t$. Here, $f_i$ is a quenched random variable which for each
particle is distributed according to a common distribution ${\mathcal P}(f)$, while
the
quantity $\Delta V(\mathcal{C})$ is the change in the potential energy due to
the attempted hop from $\th_i$ to $\th_i'$: $\Delta
V(\mathcal{C})=(1/N)\sum_{j=1}^N[-\cos(\th_i'-\th_j)+\cos(\th_i-\th_j)]+
h_i [\cos \th_i'-\cos \th_i]$.
The function $g$ is of the form $g(x)=(1/2)(1-\tanh(\beta x/2))$,
where $\beta$ is the inverse temperature. The dynamics models the
overdamped motion of the particles in contact with
an external heat bath at inverse temperature $\beta$ and in presence of
an external field. The case $p \ne q$ for which the particles move
asymmetrically forward and backward mimics the action of an
external drive that makes the particles to move in one preferential direction along the
circle. Note that in the dynamics, the initial ordering of particles on
the circle is not conserved in time. Taking $f_i$'s as quenched random
variables introduces in the dynamics a different source of noise than the one due to the Monte
Carlo update scheme which is annealed in nature.

There are two sources of quenched randomness in the model through
the presence of (i) jump lengths $f_i$'s, and (ii) field strengths
$h_i$'s. Later,
we will consider specifically the first of the two sources of
randomness, and take the $h_i$'s to be the same for all particles. In the conclusions, we will comment
on how our analytical approach may be easily adapted to consider the
randomness due to the $h_i$'s.

Let $P=P(\{\th_i\};t)$ be the probability to observe the configuration
$\mathcal{C}=\{\th_i\}$ at time $t$. In the limit of continuous time, the evolution of $P$ is given by the Master equation, which
may be derived by considering the change in $P$ in a small time $\Delta t$ according to the dynamical evolution rules given above,
and then taking the limit $\Delta t \to 0$ while keeping $f_i$'s fixed
and finite. Defining $\Delta \th_{ij}=\th_i-\th_j$, we get the Master
equation as
\bea
&&\hspace{-1.5cm}\fr{\partial P}{\partial t}=\sum_{i=1}^N
\Big[P(\ldots,\th_i-f_i,\ldots;t) \nonumber \\
&&\hspace{-1.5cm}\times p\Big\{1-\tanh\fr{\beta}{2}\Big(\fr{1}{N}\sum_{j=1}^N[-\cos
\Delta \theta_{ij}+\cos(\Delta \th_{ij}-f_i)]+
h_i [\cos \th_i-\cos (\th_i-f_i)]\Big)\Big\}\nonumber \\
&&\hspace{-1.5cm}+P(\ldots,\th_i+f_i,\ldots;t) \nonumber \\
&&\hspace{-1.5cm}\times q\Big\{1-\tanh\fr{\beta}{2}\Big(\fr{1}{N}\sum_{j=1}^N[-\cos
\Delta \th_{ij}+\cos(\Delta \th_{ij}+f_i)]+
h_i [\cos\th_i-\cos (\th_i+f_i)]\Big)\Big\}\nonumber \\
&&\hspace{-1.5cm}-P(\ldots,\th_i,\ldots;t) \nonumber \\
&&\hspace{-1.5cm}\times\Big(p\Big\{1-\tanh\fr{\beta}{2}\Big(\fr{1}{N}\sum_{j=1}^N[-\cos(\Delta
\th_{ij}+f_i)+\cos \Delta \th_{ij}+h_i [\cos(\th_i+f_i)-\cos \th_i]\Big)\Big\}\nonumber \\
&&\hspace{-1.5cm}+q\Big\{1-\tanh\fr{\beta}{2}\Big(\fr{1}{N}\sum_{j=1}^N[-\cos(\Delta
\th_{ij}-f_i)+\cos \Delta \th_{ij}]+h_i
[\cos(\th_i-f_i)-\cos \th_i]\Big)\Big\}\Big)\Big].
\l{master-eqn}
\eea

At long times, the system settles into a stationary state
corresponding to the time-independent probability $P_{\rm st}(\{\th_i\})$. For
$p=1/2$, the particles attempt to move forward and backward in a symmetric
manner, and the system has an equilibrium stationary
state in which the condition of detailed balance is satisfied with the measure $P_{\rm
eq}(\{\th_i\}) \propto
e^{-\beta V(\{\th_i\})}$. On the other hand, for $p \ne 1/2$, the
particles have a preferred direction to hop on the circle, and the system at
long times settles into a nonequilibrium stationary state, characterized
by a violation of detailed balance leading to nonzero
probability currents in phase space. In the
absence of the external field, the dynamics with $p=1/2$ samples the equilibrium
measure of the Brownian mean-field model of long-range
interacting systems, developed as an extension of the microcanonical
dynamics of the HMF model to a canonical dynamics that mimics the
interaction of the system with an external heat bath \cite{Chavanis:2011}. 

\section{Fokker-Planck limit}
\l{Langevin}
Here, we analyze the Fokker-Planck limit of the dynamics of our
model. To this end, we first obtain the Fokker-Planck
equation corresponding to the Master equation (\ref{master-eqn}) by
making the assumption that $f_i \ll 1 ~\forall~ i$, so that we may
Taylor expand functions in powers of
$f_i$'s \cite{vanKampen}. As shown in Appendix A, retaining terms to second-order in $f_i$'s, we get the Fokker-Planck
equation for $P(\{\th_i\};t)$ as
\be
\fr{\partial P}{\partial t}=-\sum_{i=1}^N \fr{\partial J_i}{\partial
\th_i},
\l{FP}
\ee
where the probability current $J_i$ for the $i$th particle is given by
\bea
J_i&=&\Big[(2p-1)f_i+
\fr{f_i^2 \beta}{2}\Big(\fr{1}{N}\sum_{j=1}^N\sin \Delta \th_{ji}+h_i\sin
\th_i\Big)\Big]P-\fr{f_i^2}{2}\fr{\partial P}{\partial
\th_i}.
\eea
The corresponding Langevin equation is easily written down as
\be
\dot{\th_i}=(2p-1)f_i+\fr{f_i^2\beta}{2}\Big(\fr{1}{N}\sum_{j=1}^N\sin(\th_j-\th_i)+h_i\sin\th_i\Big)+f_i \eta_i(t),
\l{EOM}
\ee
where the dot denotes derivative with respect to time, and $\eta_i(t)$
is a random noise with
\be
\langle \eta_i(t) \rangle=0, ~~~~\langle \eta_i(t)\eta_j(t') \rangle=\delta_{ij}\delta(t-t').
\ee

From the Fokker-Planck equation (\ref{FP}), it is evident that, as in
the finite-$f_i$ dynamics, the system for $p=1/2$ settles into the equilibrium stationary state with $P_{\rm
eq}(\{\th_i\})$ which makes $J_i=0$
individually for each $i$. On the other
hand, for $p \ne 1/2$, the system at long times reaches a
non-equilibrium stationary state. However, in the special case when the
jump length is the same for all the particles and there is no external
field ($f_i=f$ and $h_i=0 ~\forall~ i$), one may make a Galilean
transformation, $\th_i \to \th_i+[(2p-1)f/2] t$, so that in the frame moving
with the velocity $[(2p-1)f/2]$, the Langevin equation (\ref{EOM})
takes a form identical to the one for $p=1/2$, and the stationary state
has the equilibrium measure $P_{\rm
eq}(\{\th_i\})$.

\subsection{Limit $N \to \infty$ and constant field: Single-particle distribution}
In the thermodynamic limit $N \to \infty$, when the external field is the same for all
the particles, $h_i=h$, let us define the single-particle distribution $\rho(\th;f,t)$ such that $\rho(\th;f,t)$ gives the density of particles with jump length $f$
which are at location $\th$ on the circle at time $t$. We have $\rho(\th;f,t)=\rho(\th+2\pi;f,t)$, and also the
normalization
\be
\int_0^{2\pi} d\th ~\rho(\th;f,t)=1~~\forall~~ f.
\l{norm}
\ee

In terms of $\rho(\th;f,t)$, the Langevin equation (\ref{EOM}) in the
limit $N \to \infty$ for a
particle with jump length $f$ and at position $\th$ reads 
\be
\dot{\th}=(2p-1)f+\fr{f^2\beta}{2} \Big(m_y \cos
\th-m_x \sin \th+ h\sin \th\Big)+f\eta(t),
\l{EOM1}
\ee
where
\bea
&&(m_x,m_y)=\int d\th df ~(\cos \th, \sin
\th)\rho(\th;f,t)\mathcal{P}(f), 
\eea
and
\be
\langle \eta(t) \rangle=0, ~~~~\langle \eta(t)\eta(t')
\rangle=\delta(t-t').
\ee
Let us note that the dynamics (\ref{EOM1}) is similar with $\eta(t)=0$ to
that of the Kuramoto model of synchronization \cite{Kuramoto:1984} and
with $\eta(t) \ne 0$ to that of its extension considered in Ref.
\cite{Sakaguchi:1988} that includes noise. However, in the latter case, a crucial difference is that in
Eq. (\ref{EOM1}), the noise term and the drift term (the first term on the
right hand side) contain the same
factor $f$, and are therefore, related, unlike the model in Ref. \cite{Sakaguchi:1988}.

The single-particle Fokker-Planck equation satisfied by $\rho(\th;f,t)$ may be obtained from the
Langevin equation (\ref{EOM1}) as
\be
\fr{\partial \rho}{\partial t}=-\fr{\partial j}{\partial
\th},
\l{FP-sp}
\ee
where the probability current $j$ is given by
\bea
j&=&\Big[(2p-1)f+
\fr{f^2\beta}{2} \Big(m_y \cos \th-m_x \sin \th+ h\sin \th\Big)\Big]\rho-\fr{f^2}{2}\fr{\partial
\rho}{\partial \th}. 
\eea
\begin{figure}[here]
\includegraphics[width=80mm]{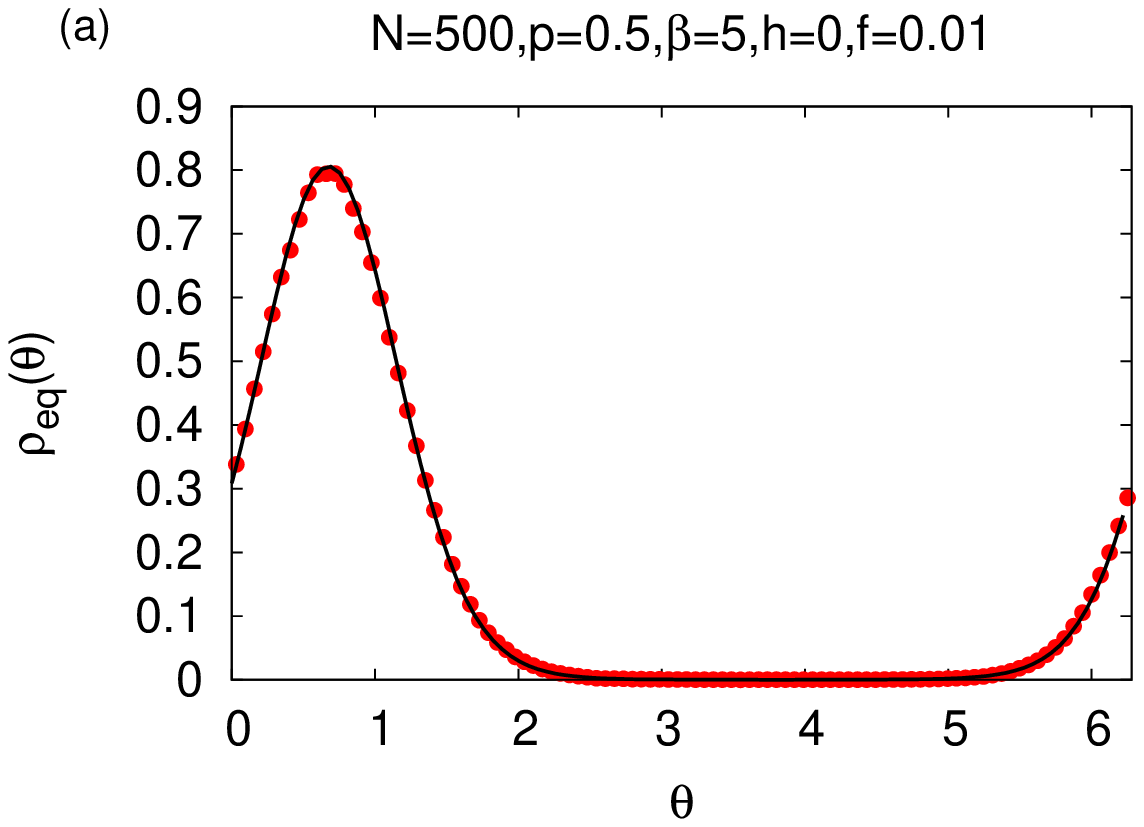}
\includegraphics[width=80mm]{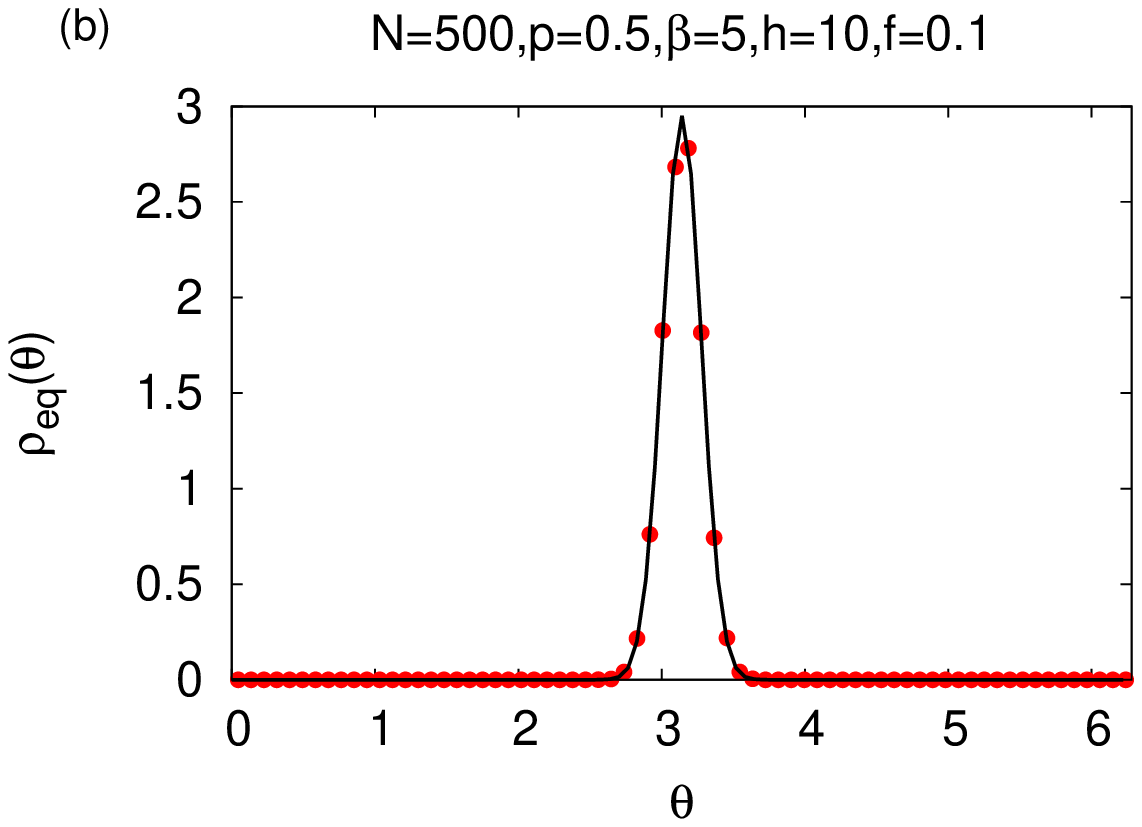}
\caption{Equilibrium distribution $\rho_{\rm eq}(\th)$: Theory (continuous line) in the
Fokker-Planck approximation and in the limit $N \to \infty$, given by
Eq. (\ref{eq-sp}), compared with Monte Carlo simulations (points),
taking $f=0.01, h=0$ (panel (a)) and $f=0.1,h=10$ (panel (b)). We observe an excellent
agreement between theory and simulations.
In (a), when there is no field, the distribution is centered around a
value of $\theta$ which is arbitrary; this corresponds to a spontaneous breaking 
of the $O(2)$ symmetry of the potential $\mathcal{V}(\mathcal{C})$.}
\label{eq-h1}
\end{figure}

The stationary solution $\rho_{\rm st}$ of the Fokker-Planck equation
(\ref{FP-sp}) is given by
 (see
Appendix B)
\bea
&&\rho_{\rm st}(\th;f)=\rho(0;f)e^{2(2p-1)\th/f+\beta (m_x\cos\th+m_y
\sin \th- h\cos \th)}\nonumber \\
&&\times\left[1+(e^{-4\pi(2p-1)/f}-1)
\fr{\displaystyle \int_0^\th d\th' e^{-2(2p-1)\th'/f-\beta (m_x \cos\th'+m_y
\sin \th'-h\cos \th')}}{\displaystyle \int_{0}^{2\pi} d\th'
e^{-2(2p-1)\th'/f-\beta(m_x \cos\th'+m_y \sin \th'- h\cos
\th')}}\right],  
\l{sp-st-soln}
\eea
where $(m_x,m_y)=\int d\th df (\cos \th,\sin \th)\rho_{\rm
st}(\th;f)\mathcal{P}(f)$, and the constant $\rho(0;f)$ is fixed by the
normalization condition (\ref{norm}).

When the jump length is the same for all particles, $f_i=f$, we have
\bea
&&\rho_{\rm st}(\th)=\rho(0)e^{2(2p-1)\th/f+\beta (m_x\cos\th+m_y
\sin \th- h\cos \th)}\nonumber \\
&&\times \left[1+(e^{-4\pi(2p-1)/f}-1)\fr{\displaystyle \int_0^\th d\th' e^{-2(2p-1)\th'/f-\beta (m_x \cos\th'+m_y
\sin \th'-h\cos \th')}}{\displaystyle \int_{0}^{2\pi} d\th'
e^{-2(2p-1)\th'/f-\beta(m_x \cos\th'+m_y \sin \th'- h\cos
\th')}}\right], 
\l{sp-st-soln-fsame}
\eea
where the constant $\rho(0)$ is fixed by normalization: $\int_0^{2\pi} d\th
~\rho_{\rm st}(\th)=1$.
For $p=1/2$, we obtain the equilibrium single-particle distribution as
\bea
&&\rho_{\rm eq}(\th)=\fr{e^{\beta (m_x\cos\th+m_y\sin \th-h\cos
\th)}}{\displaystyle \int_0^{2\th} d\th e^{\beta (m_x\cos\th+m_y\sin \th-h\cos
\th)}}.
\l{eq-sp}
\eea
It is worthwhile to point out that the
equilibrium distribution (\ref{eq-sp}) does not depend on the value of the jump length $f$, as does the nonequilibrium distribution (\ref{sp-st-soln-fsame}).

\section{Numerical simulations}
\begin{figure}[here]
\includegraphics[width=80mm]{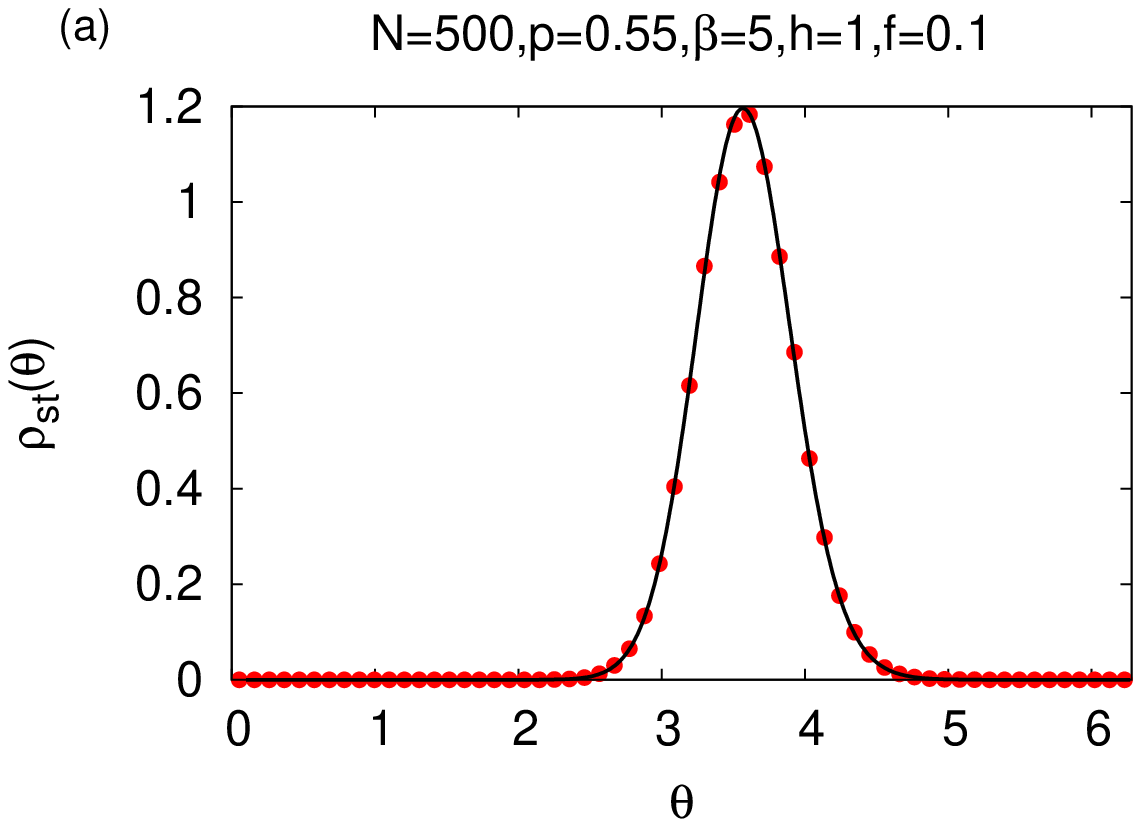}
\includegraphics[width=80mm]{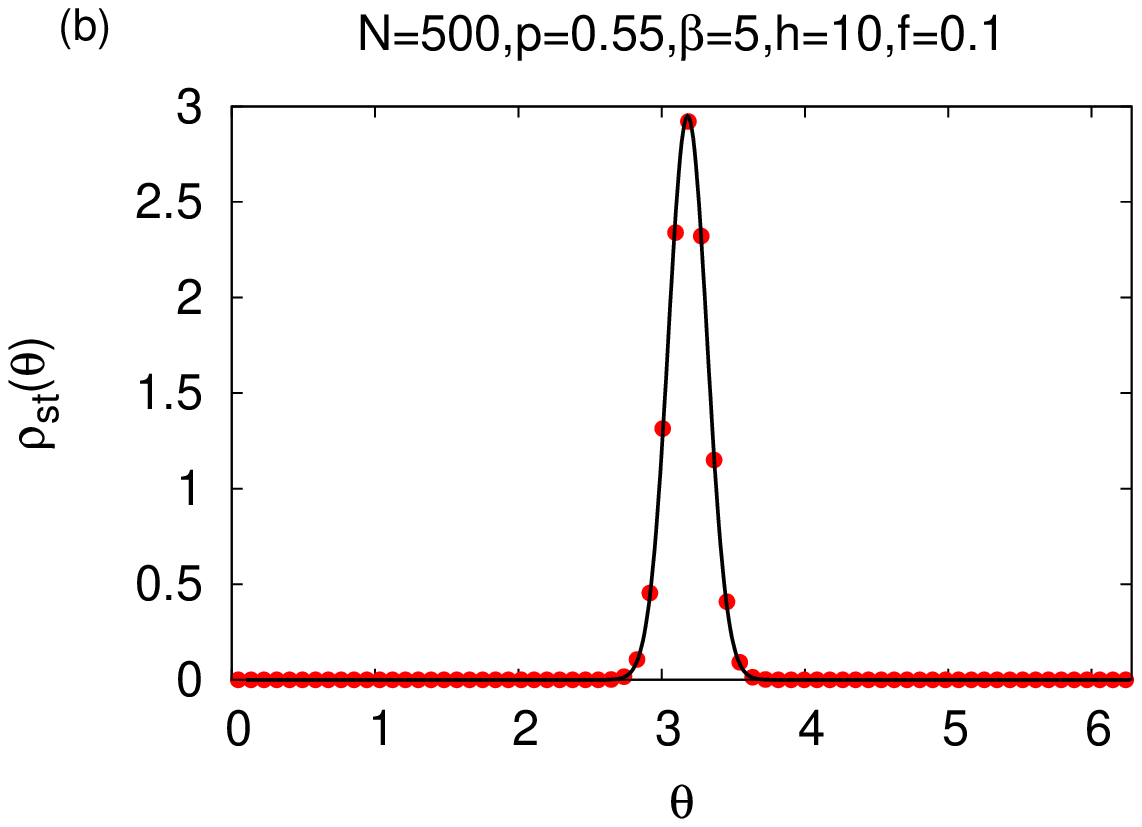}
\includegraphics[width=80mm]{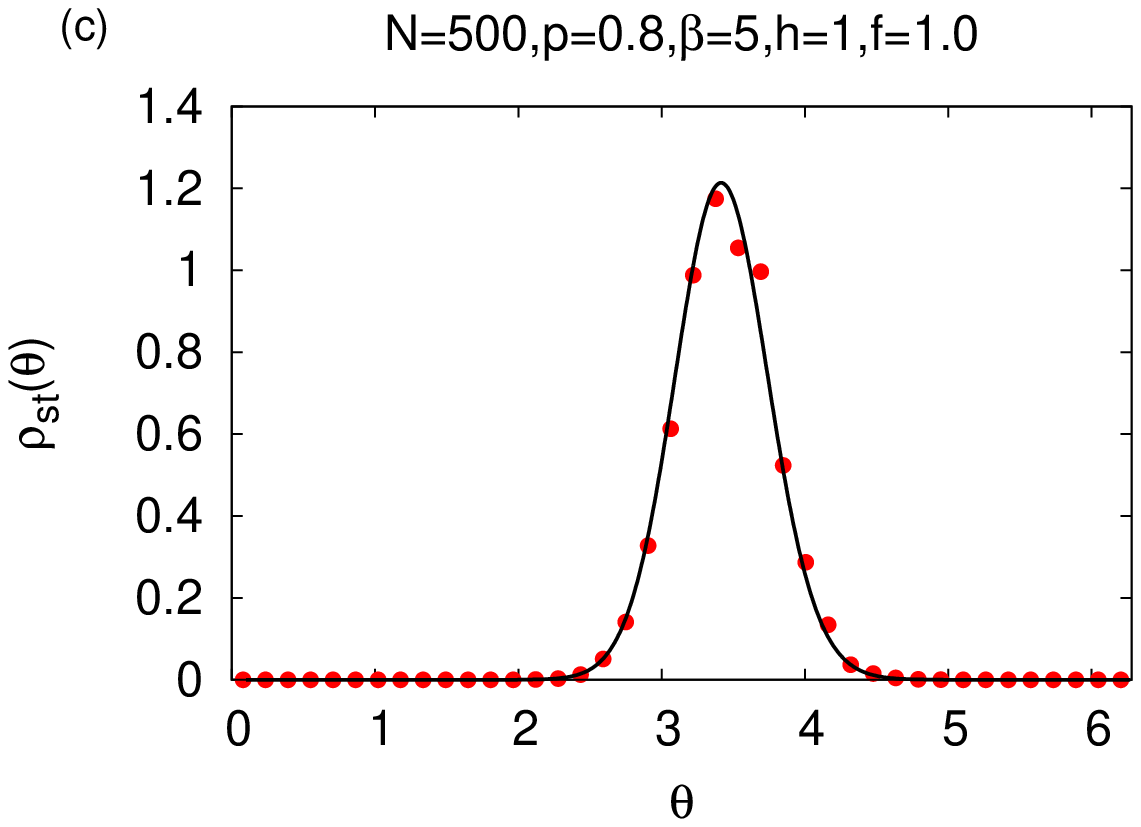}
\includegraphics[width=80mm]{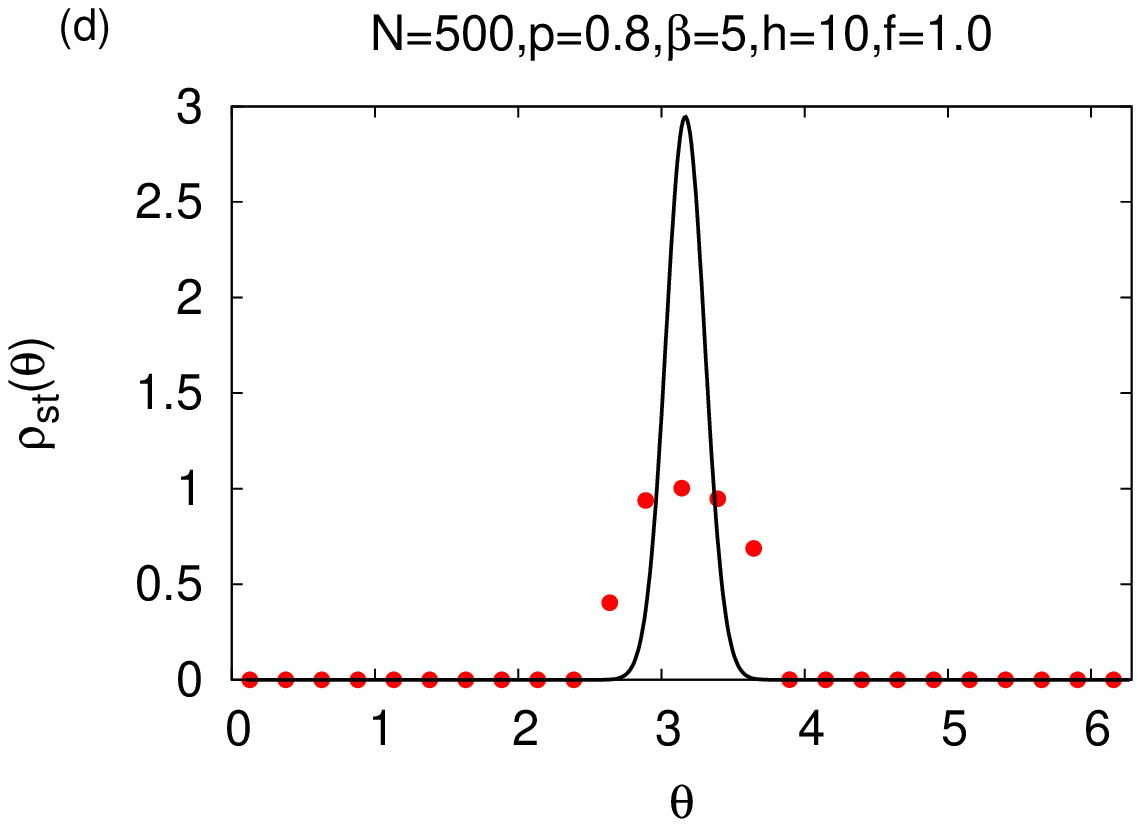}
\caption{Stationary distribution $\rho_{\rm st}(\th)$ for the case $p
\ne 1/2$ and $h \ne 0$: Panels (a) and (b) compare
Monte Carlo simulation results (points) for $f=0.1$ and theory (continuous lines) in the
Fokker-Planck approximation and in the limit $N \to \infty$, given by
Eq. (\ref{sp-st-soln-fsame}), illustrating an excellent agreement.
In panels (c) and (d) for $f=1$, we observe the expected disagreement between
simulations and the Fokker-Planck-limit theory valid for $f
\ll 1$. }
\label{st-h1}
\end{figure}

\l{numerics}
Choosing the jump length $f_i\ll 1$ to be the same for all particles, we show in Fig.
\ref{eq-h1} a comparison of the $N$-particle Monte Carlo
simulation results for $\rho_{\rm eq}(\th)$, obtained for $N=500$, and
and its theoretical form, Eq. (\ref{eq-sp}), applicable in the
Fokker-Planck approximation and in the limit $N \to \infty$. 
We observe an excellent agreement between simulations and theory. For
the case $p \ne 1/2$ and $h \ne 0$, Fig. \ref{st-h1}(a),(b) compare
simulation results and theory (Eq. (\ref{sp-st-soln-fsame})) for $\rho_{\rm st}(\th)$ for two values of $h$, again illustrating an excellent agreement.
Figures \ref{st-h1}(c),(d) compare simulation results for
$f=1$ with the Fokker-Planck-limit theory valid for $f
\ll 1$; in (c), we see a reasonable agreement, while in (d), the
disagreement is quite large. For the latter case, we have checked that
for the same parameter values, the mismatch between theory and
simulations does not reduce with larger $N$, which implies that it is due to the large value of $f$ used as compared to the Fokker-Planck limit, and not due to finiteness of $N$.

\section{Conclusions}
In this work, we introduced a model of long-range interacting systems
involving $N$ globally coupled particles moving on a circle. The system
evolves under a stochastic Monte Carlo dynamics consisting of particle jumps, either
symmetrically to the left and to the right, or, asymmetrically, by quenched
random amounts sampled from a common distribution. The attempted new
locations of the particles are accepted with transition rates chosen in
such a way that for symmetric jumps, the stationary state of the system
is the Gibbs equilibrium state. For asymmetric jumps, the system at long
times reaches a nonequilibrium steady state characterized by nonzero
probability currents in phase space. For the associated Fokker-Planck
dynamics in the thermodynamic limit $N \to \infty$, we computed exactly the steady
state distribution and found that it has a nontrivial form that reduces
to the Gibbs distribution in the equilibrium limit, see Eqs.
(\ref{sp-st-soln-fsame}) and (\ref{eq-sp}). We compared our
theoretical predictions with $N$-particle Monte Carlo simulations, and
found an excellent agreement between the two in the Fokker-Planck limit.
The observed disagreement when the limit is not satisfied opens up the
very interesting scope of analyzing and obtaining corrections to the Fokker-Planck
answer. It is also of interest to treat the external fields $h_i$'s
in Eq. (\ref{master-eqn}) as quenched random variables sampled from a common
distribution. 
It is easy to generalize our
analytical framework to treat this case in the Fokker-Planck limit by
considering instead of $\rho(\th;f,t)$ the distribution $\rho(\th;f,h,t)$
giving the density of particles with jump length $f$ and under the
action of field with strength $h$, which are at location $\th$ at time
$t$. One would then have to obtain the Fokker-Planck
equation that $\rho(\th;f,h,t)$ satisfies, by considering $f_i \ll 1$ and
$h_i \ll 1 ~\forall~ i$ in the Master equation (\ref{master-eqn}) and performing suitable Taylor series
expansion of functions of
$f_i$'s and $h_i$'s.
With quenched $h_i$'s, one may consider the canonical dynamics introduced
in this paper and its grand canonical counterpart (where, say, the number of particles $N$ is not conserved), 
and investigate the issue of equivalence of the nonequilibrium steady state under the two dynamics.
This issue is particularly relevant, since long-range interacting
systems are known to show inequivalence in equilibrium in presence of
random fields \cite{Bertalan:2011}. 
\section{Acknowledgements}
SG acknowledges the support of the CEFIPRA Grant 4604-3 and the contract
ANR-10-CEXC-010-01, and the hospitality of ENS-Lyon. We acknowledge
fruitful discussions with Sergio Ciliberto.

\section{Appendix A: Derivation of the Fokker-Planck equation (\ref{FP})}
Considering the Master equation (\ref{master-eqn}) for $f_i \ll 1 ~\forall~ i$,
we expand all functions of $f_i$'s in powers of $f_i$'s. Retaining
terms to second-order in $f_i$'s, we get 
\bea
&&\fr{\partial P}{\partial t}\approx \sum_{i=1}^N\Big[(P-f_i\fr{\partial P}{\partial
\th_i}+\fr{f_i^2}{2}\fr{\partial^2 P}{\partial
\th_i^2})p\Big\{1-\fr{\beta}{2}\Big(\fr{1}{N}\sum_{j=1}^N[-\fr{f_i^2}{2}\cos
\Delta \th_{ij}+f_i\sin \Delta\th_{ij}]\nonumber \\
&&+h_i[\fr{f_i^2}{2}\cos \th_i-f_i\sin \th_i]\Big)\Big\}\nonumber \\
&&+(P+f_i\fr{\partial P}{\partial \th_i}+\fr{f_i^2}{2}\fr{\partial^2
P}{\partial
\th_i^2})q\Big\{1-\fr{\beta}{2}\Big(\fr{1}{N}\sum_{j=1}^N[-\fr{f_i^2}{2}\cos
\Delta \th_{ij}-f_i\sin \Delta\th_{ij}]\nonumber \\
&&+h_i[\fr{f_i^2}{2}\cos \th_i+f_i\sin \th_i]\Big)\Big\}\nonumber \\
&&-Pp\Big\{1-\fr{\beta}{2}\Big(\fr{1}{2N}\sum_{j=1}^N[\fr{f_i^2}{2}\cos
\Delta \th_{ij}+f_i\sin \Delta \th_{ij}]+h_i[-\fr{f_i^2}{2}\cos \th_i-f_i \sin \th_i]\Big)\Big\}\nonumber \\
&&-Pq\Big\{1-\fr{\beta}{2}\Big(\fr{1}{N}\sum_{j=1}^N[\fr{f_i^2}{2}\cos
\Delta \th_{ij}-f_i\sin \Delta\th_{ij}]+h_i[-\fr{f_i^2}{2}\cos \th_i+f_i \sin \th_i]\Big)\Big\}\Big] \nonumber \\
&&=-\sum_{i=1}^N \fr{\partial J_i}{\partial
\th_i},
\l{FP-app}
\eea
where the probability current $J_i$ for the $i$th particle is given by
\bea
J_i&=&\Big[(2p-1)f_i+
\fr{f_i^2\beta}{2}\Big(\fr{1}{N}\sum_{j=1}^N\sin \Delta \th_{ji}+h_i\sin
\th_i\Big)\Big]P-\fr{f_i^2}{2}\fr{\partial P}{\partial
\th_i}.
\eea
\section{Appendix B: Stationary solution of Eq. (\ref{FP-sp})}
\l{FP-sp-st-soln}
Here, we obtain the steady state solution of Eq. (\ref{FP-sp}). A similar
equation and the steady state solution appear in Ref.
\cite{Stratonovich:1967}. In the steady
state, we have 
\be
\fr{\partial\rho_{\rm st}}{\partial \th}-\Big(\fr{2(2p-1)}{f}+\beta
m_{\rm st}\sin(\psi_{\rm st}-\th)+\beta h\sin \th\Big)\rho_{\rm st}=C,
\l{FP-sp-st-1}
\ee
where $C$ is a constant independent of $\th$, and we have defined
\bea
m_{\rm st}=\sqrt{m_x^2+m_y^2}; ~~\psi_{\rm st}=\tan^{-1}(m_y/m_x), \\
(m_x,m_y)=\int d\th df (\cos \th,\sin \th)\rho_{\rm
st}(\th;f)\mathcal{P}(f). 
\eea
Multiplying both sides of Eq. (\ref{FP-sp-st-1}) by
$\exp[-2(2p-1)\th/f-\beta m_{\rm st}\cos(\psi_{\rm
st}-\th)+\beta h\cos \th]$, and then integrating over $\th$, we get 
\bea
&&\rho_{\rm st}(\th;f)=\rho(0;f)e^{2(2p-1)\th/f+\beta m_{\rm
st}[\cos(\psi_{\rm st}-\th)-\cos \psi_{\rm st}]+\beta h(1-\cos \th)}\nonumber \\
&&+Ce^{2(2p-1)\th/f+\beta m_{\rm
st}\cos(\psi_{\rm st}-\th)-\beta h\cos \th}\int_0^\th d\th' e^{-2(2p-1)\th'/f-\beta m_{\rm
st}\cos(\psi_{\rm st}-\th')+\beta h\cos \th'}, \nonumber \\ 
\eea
where $\rho(0;f)=\rho(\th;f,0)$ is the initial condition at time $t=0$. 
Requiring that $\rho_{\rm st}(\th+2\pi;f)=\rho_{\rm
st}(\th;f)$ fixes $C$ to be
\be
C=\fr{\rho(0;f)e^{-\beta m_{\rm st}\cos \psi_{\rm st}+\beta h}(e^{-4\pi
(2p-1)/f}-1)}{\displaystyle\int_{0}^{2\pi} d\th' e^{-2(2p-1)\th'/f-\beta m_{\rm st}\cos(\psi_{\rm
st}-\th')+\beta h\cos \th'}},
\ee
and hence,
\bea
&&\rho_{\rm st}(\th;f)=\rho(0;f)e^{2(2p-1)\th/f+\beta m_{\rm
st}[\cos(\psi_{\rm st}-\th)-\cos \psi_{\rm st}]+\beta h(1-\cos \th)}\nonumber \\
&&\times\left[1+(e^{-4\pi(2p-1)/f}-1)\fr{\displaystyle \int_0^\th d\th' e^{-2(2p-1)\th'/f-\beta m_{\rm
st}\cos(\psi_{\rm st}-\th')+\beta h\cos \th'}}{\displaystyle \int_{0}^{2\pi} d\th'
e^{-2(2p-1)\th'/f-\beta m_{\rm st}\cos(\psi_{\rm
st}-\th')+\beta h\cos \th'}}\right].  
\eea
Redefining $\rho(0;f)$, and reverting to the variables $m_x$ and $m_y$, we get
\bea
&&\rho_{\rm st}(\th;f)=\rho(0;f)e^{2(2p-1)\th/f+\beta (m_x\cos\th+m_y
\sin \th- h\cos \th)}\nonumber \\
&&\times \left[1+(e^{-4\pi(2p-1)/f}-1) \fr{\displaystyle \int_0^\th d\th' e^{-2(2p-1)\th'/f-\beta (m_x \cos\th'+m_y
\sin \th'-h\cos \th')}}{\displaystyle \int_{0}^{2\pi} d\th'
e^{-2(2p-1)\th'/f-\beta(m_x \cos\th'+m_y \sin \th'- h\cos
\th')}}\right],
\eea
where $\rho(0;f)$ is fixed by the normalization: $\int_0^{2\pi} d\th ~\rho_{\rm st}(\th;f)=1$.

\end{document}